\documentclass[onecolumn,peerreview]{IEEEtran}
\ifCLASSINFOpdf
\else
    \usepackage[dvips]{graphicx}
\fi
\usepackage{amsmath,amssymb}
\newcommand{\bysame}{%
    \leavevmode\hbox to 3em{\hrulefill}\,}

\begin{document}
%
\title{Non-Linear Estimation of Convolutionally Encoded Sequences}
%
%
%

\author{Masato~Tajima,~\IEEEmembership{Senior~Member,~IEEE}
\thanks{M. Tajima is with University of Toyama, 3190 Gofuku,
Toyama 930-8555, Japan (e-mail: masatotjm@kind.ocn.ne.jp).}
\thanks{Manuscript received April 19, 2005; revised August 26, 2015.}}

%
%

\markboth{Journal of \LaTeX\ Class Files,~Vol.~14, No.~8, August~2015}%
{Tajima \MakeLowercase{\textit{}}: Non-Linear Estimation of Convolutionally Encoded Sequences}
%



\maketitle

\begin{abstract}
Suppose that a convolutionally encoded sequence is transmitted symbol by symbol over an AWGN channel using BPSK modulation. In this case, pairs of the signal (i.e., code symbol) and observation are not jointly Gaussian and therefore, a linear estimation method cannot be applied. Hence, in this paper, non-linear estimation of convolutionally encoded sequences is discussed. First a probability measure (denoted $Q$), whose Radon-Nikodym derivative with respect to the underlying probability measure $P$ is an exponential martingale, is constructed. It is shown that with respect to $Q$, the observations are mutually independent Gaussian random vectors with zero mean and identity covariance matrix. We see that the relationship between observation noises (with respect to $P$) and observations (with respect to $Q$) has a close relation to the Girsanov theorem in continuous case. Next, using the probability measure $Q$, we calculate the conditional probability of an event related to any encoded symbol conditioned by the observations. Moreover, we transform it into a recursive form. In the process of derivation, the metric associated with an encoded sequence comes out in a natural way. Finally, it is shown that maximum {\it a posteriori} probability (MAP) decoding of convolutional codes is realized using the derived conditional probability.
\end{abstract}

\begin{IEEEkeywords}
Convolutional codes, non-linear estimation, Radon-Nikodym derivative, Girsanov theorem, MAP decoding.
\end{IEEEkeywords}

%
\IEEEpeerreviewmaketitle

\section{Introduction}
%
%
%
%
Consider an error control coding scheme. A decoder estimates a transmitted message based on the received noisy data. Hence, it is natural to think that error control coding has a close connection with estimation of stochastic processes. In this paper, we consider convolutional coding/decoding from the viewpoint of the filtering (or smoothing) theory for discrete-time stochastic processes (cf.~\cite{taji 19}). To begin with, we state some basic notions needed in this paper. In the following, the underlying probability space $(\Omega, \mathcal{F}, P)$ is implicitly assumed. Here, $\mathcal{F}$ is a $\sigma$-field of subsets of $\Omega$, and $P$ is a probability measure defined on $\mathcal{F}$. Let $X$ be a real-valued random variable defined on $\Omega$ (the set of real numbers is denoted by $R$). In this paper, random variables are expressed in capital letters in principle. Denote by $\sigma(X)$ the $\sigma$-field generated by $X$. Let $\{X_n,~n \in N\}$ ($N$ denotes the set of natural numbers) be a family of real random variables. Then the smallest $\sigma$-field which contains $\cup_{n \in N}\sigma(X_n)$ is denoted by $\sigma(X_n,~n \in N)$ (or $\vee_{n \in N}\sigma(X_n)$). Let $\mathcal{B}$ be a sub-$\sigma$-field of $\mathcal{F}$. The conditional expectation of $X$ with respect to $\mathcal{B}$ is denoted by $E(X \vert \mathcal{B})$ ($E(\cdot)$ is the expectation).
\par
First assume the following:
\begin{itemize}
\item [1)] $W_k~(k \in N)$ are mutually independent {\it Gaussian} random vectors of dimension $d$ with mean {\boldmath $0$} and covariance matrix $I_d$ ($I_d$ is the identity matrix of size $d \times d$). ($W_k$ represents an observation noise.)
\item [2)] $X_k~(k \in N)$ are {\it Gaussian} random vectors of dimension $s$, and they are independent of $W_k~(k \in N)$. ($X_k$ represents a signal.)
\item [3)] Observations $Z_k$ are given by
\begin{equation}
Z_k=X_kC_k+W_k,~~k \geq 1 ,
\end{equation}
where $C_k$ is an $s \times d$ matrix.
\end{itemize}
Under these conditions, we see that
\begin{displaymath}
\{X_k^{(i)}, W_l^{(j)}; 1 \leq i \leq s, 1 \leq j \leq d,~k, l \in N\}
\end{displaymath}
forms a system of Gaussian random variables (i.e., every linear combination of variables contained in the set has a Gaussian distribution). Moreover, we see that $(X_k, Z_k)$ are ``jointly Gaussian''. (Hence, a ``linear'' estimation method can be applied.) Denote by $\mathcal{B}_n=\sigma(Z_1, \cdots, Z_n)$ the $\sigma$-field which represents the observations obtained up to time $n$. Also, Let $H_n$ be the Gaussian space~\cite{hida 80,ito 91,kuni 76} generated by $Z_1, \cdots, Z_n$. Then the best estimate (i.e., the least-squares estimate) for $X_n^{(i)}~(1 \leq i \leq s)$ based on $\mathcal{B}_n$ (denoted by $\hat X_n^{(i)}$) is given by the conditional expectation $E(X_n^{(i)} \vert \mathcal{B}_n)$. Note that under above conditions, we have $E(X_n^{(i)} \vert \mathcal{B}_n)=P_{H_n}X_n^{(i)}$~\cite{hida 80,ito 91,kuni 76,oksen 98}, where $P_{H_n}X_n^{(i)}$ is the orthogonal projection of $X_n^{(i)}$ onto the space $H_n$.
\par
{\it Remark:} Let $\mathcal{B} (\subset \mathcal{F})$ be a sub-$\sigma$-field. Also, let $X \in L^2$ be $\mathcal{F}$-measurable, where $L^2$ is the set of random variables such that $E(\vert X \vert^2)<\infty $. Let us define as $L^2(\mathcal{B})\stackrel{\triangle}{=}\{Y \in L^2; Y~\mbox{is}~\mathcal{B}-\mbox{measurable}\}$. Denote by $P_{L^2(\mathcal{B})}$ the orthogonal projection from $L^2$ onto the sub-space $L^2(\mathcal{B})$. Then~\cite{ito 91,kuni 76,nev 65,oksen 98,will 91} we have
\begin{equation}
P_{L^2(\mathcal{B})}X=E(X \vert \mathcal{B}) .
\end{equation}
\par
Since $P_{L^2(\mathcal{B}_n)}X_n^{(i)}=E(X_n^{(i)} \vert \mathcal{B}_n)$ holds (see the above remark), it follows that
\begin{equation}
P_{L^2(\mathcal{B}_n)}X_n^{(i)}=P_{H_n}X_n^{(i)} .
\end{equation}
Note that the space $H_n$ is smaller than the space $L^2(\mathcal{B}_n)$~\cite{hida 80}. Hence, this is a remarkable feature of a system of Gaussian random variables. Furthermore, we see that $X_n^{(i)}-E(X_n^{(i)} \vert \mathcal{B}_{n-1})$ is contained in $H_n$ and is orthogonal to $H_{n-1}$. Another important feature is a close connection with the notion of {\it innovations}~\cite{kai 681,kai 98,kuni 76,oksen 98,wong 73}. In fact, it is shown that
\begin{equation}
Z_n-E(X_n \vert \mathcal{B}_{n-1})C_n,~n \geq 1
\end{equation}
are innovations associated with the observations $Z_n$~\cite{kai 681,kuni 76}. Using these properties, the well-known Kalman-Bucy filter~\cite{ari 77,jaz 07,kai 681,kuni 76,oksen 98,wong 71} is derived. The detailed derivation along the above argument is found in~\cite[Kunita]{kuni 76} and~\cite[{\O}ksendal]{oksen 98} (In the latter, a continuous-time case is dealt with). 
\par
Next, consider convolutional coding/decoding. In order to state the problem more precisely, we introduce some additional notions needed for this paper. We always assume that the underlying field is $\mbox{GF}(2)$. Let $G(D)$ be a generator matrix for an $(n_0, k_0)$ convolutional code, where $G(D)$ is assumed to be {\it canonical}~\cite{joha 99} (i.e., {\it minimal}~\cite{forn 70}). Denote by $\mbox{\boldmath $i$}=\{\mbox{\boldmath $i$}_k\}$ and $\mbox{\boldmath $y$}=\{\mbox{\boldmath $y$}_k\}$ an information sequence and the corresponding encoded sequence, respectively, where $\mbox{\boldmath $i$}_k=(i_k^{(1)}, \cdots, i_k^{(k_0)})$ is the information block at $t=k$ and $\mbox{\boldmath $y$}_k=(y_k^{(1)}, \cdots, y_k^{(n_0)})$ is the encoded block at $t=k$. In this paper, it is assumed that an encoded sequence {\boldmath $y$} is transmitted symbol by symbol over a memoryless additive white Gaussian noise (AWGN) channel using binary phase shift keying (BPSK) modulation~\cite{hell 71}. Let $\mbox{\boldmath $z$}=\{\mbox{\boldmath $z$}_k\}$ be a received sequence, where $\mbox{\boldmath $z$}_k=(z_k^{(1)}, \cdots, z_k^{(n_0)})$ is the received block at $t=k$. Each component $z_j$ of {\boldmath $z$} is modeled as
\begin{eqnarray}
z_j &=& x_j\sqrt{2E_s/N_0}+w_j \\
&=& cx_j+w_j~~(c\stackrel{\triangle}{=}\sqrt{2E_s/N_0}) .
\end{eqnarray}
Here, $x_j$ takes $\pm 1$ depending on whether the encoded symbol $y_j$ is $0$ or $1$. $E_s$ and $N_0$ denote the energy per channel symbol and the single-sided noise spectral density, respectively. Also, $w_j$ is a zero-mean unit variance Gaussian random variable. Each $w_j$ is independent of all others. 
\par
By grouping $z_j$ together as a branch, we can rewrite the observations as
\begin{displaymath}
Z_k=cX_k+W_k .
\end{displaymath}
Note that $X_k~(\in \{-1, +1\}^{n_0})$ are not Gaussian and accordingly, $Z_k$ are not Gaussian random vectors. That is, $(X_k, Z_k)$ are ``not'' jointly Gaussian.  Hence, a linear estimation method~\cite{ari 77,bala 73,kai 681,oksen 98,wong 71} cannot be applied to our case. As a result, in this paper, we will discuss ``non-linear'' estimation~\cite{fuji 72, jaz 07,kuni 76} of convolutionally encoded sequences.
\par
In order to clarify the subsequent argument, we describe the observation model which will be discussed in this paper again. Observations are given as follows:
\begin{equation}
Z_k=cX_k+W_k ,
\end{equation}
where
\begin{eqnarray}
W_k &=& \left(W_k^{(1)}, \cdots, W_k^{(n_0)}\right) \\
X_k &=& \left(X_k^{(1)}, \cdots, X_k^{(n_0)}\right) \\
Z_k &=& \left(Z_k^{(1)}, \cdots, Z_k^{(n_0)}\right) .
\end{eqnarray}
\par
Note that the following hold with respect to the triplet $(W_k, X_k, Z_k)~(k \in N)$.
\begin{itemize}
\item [1)] $W_k~(k \in N)$ are mutually independent Gaussian random vectors of dimension $n_0$ with mean {\boldmath $0$} and covariance matrix $I_{n_0}$.
\item [2)] $X_k~(k \in N)$ and $W_k~(k \in N)$ are mutually independent.
\item [3)] $Z_k~(k \in N)$ are random vectors of dimension $n_0$ and have the form $Z_k=\phi_k(X_k)+W_k$, where $\phi_k(\cdot)$ may depend on $X_j, Z_j~(j \leq k-1)$ as well. ($Z_k$ are not necessarily Gaussian.)
\end{itemize}
\par
In~\cite{kuni 76}, non-linear filtering of stochastic processes is discussed under the conditions 1), 2), and 3). Then we thought its argument can be used in our case. Hence, we will follow Kunita~\cite{kuni 76} and repeat the argument there. Also, as in~\cite{kuni 76}, we focus our attention on a conditional probability of the form $P(X_l \in B \vert \mathcal{B}_n)$, where $\mathcal{B}_n=\sigma(Z_1, \cdots, Z_n)$. In fact, using a conditional probability $P(X_l^{(i)}=x \vert \mathcal{B}_n)$,
a conditional expectation $E(X_l^{(i)} \vert \mathcal{B}_n)$ is calculated as
\begin{equation}
E(X_l^{(i)} \vert \mathcal{B}_n)=\sum_{x \in \{-1, +1\}}xP(X_l^{(i)}=x \vert \mathcal{B}_n) .
\end{equation}
\par
Now the argument in~\cite{kuni 76} is not intended to apply to the coding theory. On the other hand, our aim is convolutional coding/decoding. Hence, it is modified to meet our purpose. As a result, although proofs of the results in Section II-A have been given in~\cite{kuni 76}, we will give them again because of our modifications. Subsequently, we will derive a general conditional probability $P(X_l \in A \vert \mathcal{B}_n)~(l \leq n)$. When $l=n$, it is corresponding to {\it filtering} of $X_n$ based on $\mathcal{B}_n$, whereas when $l<n$, it is corresponding to {\it smoothing}~\cite{jaz 07,kai 682} of $X_l$ based on $\mathcal{B}_n$. In addition, we transform the obtained conditional probability into a recursive form using the structure of a code trellis. It is shown that the derived result can be used for maximum {\it a posteriori} probability (MAP) decoding~\cite{bahl 74,lin 04} of convolutional codes.
\par
The rest of the paper is organized as follows. In Section II, a new probability measure $Q$ is constructed from the original probability measure $P$ using an (exponential) martingale. Then it is shown that with respect to $Q$, the (original) observations are mutually independent Gaussian random vectors with zero mean and identity covariance matrix. We see that the relationship between observation noises (with respect to $P$) and observations (with respect to $Q$) has a close connection with the {\it Girsanov theorem}~\cite{gir 60,kai 702,kara 91,kuni 76,oksen 98} in continuous case. Next, in Section III, using the results in Section II, the conditional probability of an event related to any encoded symbol conditioned by the observations is calculated. In the process of derivation, the metric associated with an encoded sequence comes out in a natural way and we find that the argument in this paper has been connected to convolutional coding/decoding. The derived conditional probability is further transformed into a recursive form using the Markov property of state transitions on the associated code trellis. Also, the corresponding computational complexity is evaluated. Moreover, it is shown that MAP decoding~\cite{bahl 74,lin 04} of convolutional codes is realized using the obtained result. Finally, conclusions are drawn in Section IV.

\section{Exponential Martingales and Associated Probability Measures}
\subsection{Martingale $\alpha_n$ and the Associated Probability Measure $Q$}
The following lemma~\cite{ito 91, kuni 76} will be used repeatedly in our discussion.
\newtheorem{lem}{Lemma}[section]
\begin{lem}
Let $X$ and $Y$ be mutually independent random vectors of dimensions $d_1$ and $d_2$, respectively. Also, let $u(x, y)~(x \in R^{d_1}, y \in R^{d_2})$ be a bounded measurable function. Then we have
\begin{equation}
E(u(X, Y) \vert \sigma(Y))=\int u(x, Y)\mu_X(dx) ,
\end{equation}
where $\mu_X(x)$ denotes the distribution of $X$.
\end{lem}
\begin{IEEEproof}
See Appendix A.
\end{IEEEproof}
\par
{\it Remark 1:} The above lemma can be extended to a general product space. In Section III, $R^{d_1}$ is replaced with $(\{-1, +1\}^{n_0})^n$, where $n$ is a positive integer.
\par
Define as
\begin{equation}
\mathcal{F}_n\stackrel{\triangle}{=}\sigma(X_k; k \in N)\vee \sigma(W_k; k \leq n)~~(\mathcal{F}_0=\sigma(X_k; k \in N))
\end{equation}
\begin{equation}
\alpha_n\stackrel{\triangle}{=}\exp\left\{-c\sum_{k=1}^n(X_k, W_k)-\frac{1}{2}c^2n_0n \right\}~~(\alpha_0=1) .
\end{equation}
Here, $(\mbox{\boldmath $a$}, \mbox{\boldmath $b$})$ denotes the inner product of vectors {\boldmath $a$} and {\boldmath $b$}. (Define $\vert \mbox{\boldmath $a$} \vert=(\mbox{\boldmath $a$}, \mbox{\boldmath $a$})^{\frac{1}{2}}$.)
\par
Let $\mbox{\boldmath $\mathcal{G}$}=\{\mathcal{G}_n\}$ be an increasing family of sub-$\sigma$-fields of $\mathcal{F}$. Let $\{\zeta_n,~n \in N\}$ be a discrete-time stochastic process, where $\zeta_n$ is $\mathcal{G}_n$-measurable. If $E(\zeta_{n+1} \vert \mathcal{G}_n)=\zeta_n$ holds for $n \in N$, then $\{\zeta_n,~n \in N\}$ is said to be a $\mbox{\boldmath $\mathcal{G}$}$-martingale~\cite{bala 73,hida 80,ito 91,kai 98,kuni 76,nev 65,will 91,wong 71,wong 73}. We have the following.
\begin{lem}[Kunita~\cite{kuni 76}]
$\{\alpha_n\}$ is a positive $\mbox{\boldmath $\mathcal{F}$}$-martingale with mean $1$, where $\mbox{\boldmath $\mathcal{F}$}=\{\mathcal{F}_n\}$.
\end{lem}
\begin{IEEEproof}
Let us show that
\begin{displaymath}
E(\exp\{-c(X_n, W_n)-\frac{1}{2}c^2n_0\} \vert \mathcal{F}_{n-1})=1 .
\end{displaymath}
Note that $W_n$ and $\sigma(W_k; k \leq n-1)$ are mutually independent given $\sigma(X_k; k \in N)$. Hence, it follows from Lemma 2.1 that
\begin{eqnarray}
\lefteqn{E(\exp\{-c(X_n, W_n)-\frac{1}{2}c^2n_0\} \vert \mathcal{F}_{n-1})} \nonumber \\
&=& \frac{1}{(2 \pi)^{n_0/2}}\int_{R^{n_0}}\exp\{-c(X_n, y)-\frac{1}{2}c^2n_0\}\exp\{-\frac{1}{2}(y, y)\}dy \nonumber \\
&=& \frac{1}{(2 \pi)^{n_0/2}}\int_{R^{n_0}}\exp\{-\frac{1}{2}(y+cX_n, y+cX_n)\}dy=1 .
\end{eqnarray}
Multiplying both sides by $\exp\{-c\sum_{k=1}^{n-1}(X_k, W_k)-\frac{1}{2}c^2n_0(n-1)\}$ (this is $\mathcal{F}_{n-1}$-measurable), we have
\begin{displaymath}
E(\alpha_n \vert \mathcal{F}_{n-1})=\alpha_{n-1} .
\end{displaymath}
$E(\alpha_n)=1$ follows from $\alpha_0=1$ and from the property of a martingale (i.e., it has a constant mean).
\end{IEEEproof}
\par
{\it Remark 2:} From the definition of $\mathcal{F}_n=\sigma(X_k; k \in N)\vee \sigma(W_k; k \leq n)$, $\mathcal{F}_n$ contains $\sigma(X_k; k \in N)$. Accordingly, given $\sigma(X_k; k \in N)$, we can rewrite $\alpha_n$ as
\begin{displaymath}
\alpha_n=\exp\left\{-c\sum_{k=1}^n(\cdot, W_k)-\frac{1}{2}c^2n_0n \right\} .
\end{displaymath}
Then $\alpha_n$ becomes a martingale from the property of $W_k$.
\par
Now let $n_p$ be a positive integer and define a probability measure $Q$ by
\begin{equation}
Q(A)\stackrel{\triangle}{=}\int_A\alpha_{n_p}dP~~(A \in \mathcal{F}_{n_p}) .
\end{equation}
We have the following.
\begin{lem}[Kunita~\cite{kuni 76}]
Suppose that $n \leq n_p$ and $A \in \mathcal{F}_n$. Then
\begin{equation}
Q(A)=\int_A\alpha_ndP~~(A \in \mathcal{F}_n)
\end{equation}
holds.
\end{lem}
\begin{IEEEproof}
Since $\{\alpha_n\}$ is an $\mbox{\boldmath $\mathcal{F}$}$-martingale,
\begin{displaymath}
E(\alpha_{n_p} \vert \mathcal{F}_n)=\alpha_n
\end{displaymath}
holds and we have
\begin{displaymath}
\int_AE(\alpha_{n_p} \vert \mathcal{F}_n)dP=\int_A\alpha_ndP
\end{displaymath}
for $A \in \mathcal{F}_n$. Since $A \in \mathcal{F}_n$, the left-hand side is equal to $\int_A\alpha_{n_p}dP$ from the definition of a conditional expectation. Hence,
\begin{displaymath}
\int_A\alpha_{n_p}dP=\int_A\alpha_ndP~~(A \in \mathcal{F}_n)
\end{displaymath}
holds. On the other hand, if $A \in \mathcal{F}_n$, then $A \in \mathcal{F}_{n_p}$ and we have
\begin{displaymath}
Q(A)=\int_A\alpha_{n_p}dP=\int_A\alpha_ndP .
\end{displaymath}
\end{IEEEproof}
The above means that on the $\sigma$-field $\mathcal{F}_n$, $\alpha_n$ is the Radon-Nikodym derivative~\cite{hew 65,nev 65,will 91,wong 71} of $Q$ with respect to $P$ (denoted by $dQ/dP=\alpha_n$).
\par
When $X_k, Z_k~(k \in N)$ are viewed from the probability measure $Q$, we have the following.
\newtheorem{pro}{Proposition}[section]
\begin{pro}[Kunita~\cite{kuni 76}] $X_k$ and $Z_k$ have the following properties with respect to $Q$.
\begin{itemize}
\item [1)] The distribution of $X_k~(k \in N)$ with respect to $Q$ is identical to the distribution with respect to $P$.
\item [2)] $Z_k~(1 \leq k \leq n_p)$ are mutually independent Gaussian random vectors of dimension $n_0$ with mean {\boldmath $0$} and covariance matrix $I_{n_0}$.
\item [3)] $X_k~(k \in N)$ and $Z_k~(1 \leq k \leq n_p)$ are mutually independent.
\end{itemize}
\end{pro}
\begin{IEEEproof}
See Appendix B.
\end{IEEEproof}

\subsection{In Relation to the Girsanov Theorem~\cite{gir 60}}
Note Lemma 2.3 and Proposition 2.1. The probability measure $P$ is the underlying probability measure and the probability measure $Q$ is constructed from $P$ using the exponential martingale $\alpha_n$. Here reverse the positions of $P$ and $Q$. This is possible as follows. First we can assume that $Q$ is a given probability measure with respect to which conditions 2) and 3) in Proposition 2.1 are satisfied. Next, let us show that $P$ is absolutely continuous with respect to $Q$. This follows from the relation
\begin{displaymath}
Q(A)=\int_A\alpha_ndP~~(A \in \mathcal{F}_n) .
\end{displaymath}
Since $\alpha_n>0$, if $Q(A)=0$, then we have $P(A)=0$. That is, $P$ is absolutely continuous with respect to $Q$.
\par
On the other hand, we know that $dQ/dP=\alpha_n$. Hence, we have $dP/dQ=\alpha_n^{-1}$. $\alpha_n^{-1}$ is calculated as follows:
\begin{eqnarray}
\alpha_n^{-1} &=& \exp\left\{c\sum_{k=1}^n(X_k, W_k)+\frac{1}{2}c^2n_0n \right\} \nonumber \\
&=& \exp\left\{c\sum_{k=1}^n(X_k, Z_k-cX_k)+\frac{1}{2}c^2n_0n \right\} \nonumber \\
&=& \exp\left\{c\sum_{k=1}^n(X_k, Z_k)-c^2n_0n+\frac{1}{2}c^2n_0n \right\} \nonumber \\
&=& \exp\left\{c\sum_{k=1}^n(X_k, Z_k)-\frac{1}{2}c^2n_0n \right\}\stackrel{\triangle}{=}\beta_n .
\end{eqnarray}
Thus $dP/dQ=\alpha_n^{-1}=\beta_n$. Note that $\beta_n$ has an alternative expression:
\begin{equation}
\beta_n=\exp\left\{c\sum_{k=1}^n(X_k, Z_k)-\frac{1}{2}\sum_{k=1}^n(cX_k, cX_k) \right\} .
\end{equation}
From the assumption, $Z_k$ are mutually independent Gaussian random vectors of dimension $n_0$ with mean {\boldmath $0$} and covariance matrix $I_{n_0}$. Using this property, it is shown that $\{\beta_n\}$ is an $\mbox{\boldmath $\mathcal{F}$}$-martingale. Furthermore, with respect to $P$ ($dP=\beta_ndQ$),
\begin{equation}
W_k=Z_k-cX_k,~k \geq 1
\end{equation}
are mutually independent Gaussian random vectors of dimension $n_0$ with mean {\boldmath $0$} and covariance matrix $I_{n_0}$. In the above expression, the term ``$-cX_k$'' is regarded as a shift due to the change of probability measures. In summary,
\begin{itemize}
\item [1)] $Z_k$ are mutually independent Gaussian random vectors of dimension $n_0$ with mean {\boldmath $0$} and covariance matrix $I_{n_0}$ with respect to $Q$.
\item [2)] $W_k=Z_k-cX_k$ are mutually independent Gaussian random vectors of dimension $n_0$ with mean {\boldmath $0$} and covariance matrix $I_{n_0}$ with respect to $P$, where $dP=\beta_ndQ$.
\end{itemize}
We remark that $Z_k$ and $W_k$ are corresponding to the Wiener process in continuous case. Hence, we see that the above is corresponding to a discrete-time version~\cite{kara 91} of the {\it Girsanov theorem}~\cite{gir 60,kai 702,kuni 76,oksen 98} for continuous processes.

\section{Non-Linear Estimation of Convolutionally Encoded Sequences}
\subsection{Expectation Operator $E_Q$}
We have the following~\cite{kuni 76,oksen 98}.
\begin{lem} Let $P$ and $Q$ be the two probability measures defined on $\mathcal{F}$. Suppose that $Q(P)$ is absolutely continuous with respect to $P(Q)$ and let $dQ/dP=\alpha$ be the corresponding Radon-Nikodym derivative. Also, let $\mathcal{B}$ be a sub-$\sigma$-field of $\mathcal{F}$. Denote by $E_Q(\cdot)$ the expectation with respect to $Q$. Then we have
\begin{equation}
P(A \vert \mathcal{B})=\frac{E_Q(\alpha^{-1}\chi_A \vert \mathcal{B})}{E_Q(\alpha^{-1}\vert \mathcal{B})}~~(A \in \mathcal{F}) ,
\end{equation}
where $\chi_A$ is the indicator function of a set $A$.
\end{lem}
\begin{IEEEproof}
Let $A \in \mathcal{F}$ and $B \in \mathcal{B}$. Then
\begin{displaymath}
P(A \cap B)=E(\chi_A\chi_B)
\end{displaymath}
holds. On the other hand, since $dQ=\alpha dP$, we have $dP=\alpha^{-1}dQ$. Hence, the right-hand is equal to $E_Q(\alpha^{-1}\chi_A\chi_B)$. Using the properties of a conditional expectation~\cite{nev 65,will 91,wong 71}, this is modified as follows:
\begin{eqnarray}
E_Q(\alpha^{-1}\chi_A\chi_B) &=& E_Q(E_Q(\alpha^{-1}\chi_A\chi_B \vert \mathcal{B})) \nonumber \\
&=& E_Q(E_Q(\alpha^{-1}\chi_A \vert \mathcal{B})\chi_B) \nonumber \\
&=& E(\alpha E_Q(\alpha^{-1}\chi_A \vert \mathcal{B})\chi_B) \nonumber \\
&=& E(E(\alpha E_Q(\alpha^{-1}\chi_A \vert \mathcal{B})\chi_B \vert\mathcal{B})) \nonumber \\
&=& E(E(\alpha \vert \mathcal{B})E_Q(\alpha^{-1}\chi_A \vert \mathcal{B})\chi_B) .
\end{eqnarray}
Since $B \in \mathcal{B}$ is arbitrary, the above equality implies that
\begin{equation}
\chi_A=E(\alpha \vert \mathcal{B})E_Q(\alpha^{-1}\chi_A \vert \mathcal{B}) .
\end{equation}
Here note the definition of a conditional probability:
\begin{displaymath}
P(A \vert \mathcal{B})=E(\chi_A \vert \mathcal{B}) .
\end{displaymath}
By replacing $\chi_A$ by the above expression, we have $P(A \vert \mathcal{B})=E(E(\alpha \vert \mathcal{B})E_Q(\alpha^{-1}\chi_A \vert \mathcal{B}) \vert \mathcal{B})$. Since both $E(\alpha \vert \mathcal{B})$ and $E_Q(\alpha^{-1}\chi_A \vert \mathcal{B})$ are $\mathcal{B}$-measurable, these terms are put out of the expectation and we have
\begin{equation}
P(A \vert \mathcal{B})=E(\alpha \vert \mathcal{B})E_Q(\alpha^{-1}\chi_A \vert \mathcal{B}) .
\end{equation}
In particular, letting $A=\Omega$,
\begin{equation}
E(\alpha \vert \mathcal{B})=\frac{1}{E_Q(\alpha^{-1} \vert \mathcal{B})}
\end{equation}
is obtained. Then by substituting $\frac{1}{E_Q(\alpha^{-1} \vert \mathcal{B})}$ for $E(\alpha \vert \mathcal{B})$, we have
\begin{displaymath}
P(A \vert \mathcal{B})=\frac{E_Q(\alpha^{-1}\chi_A \vert \mathcal{B})}{E_Q(\alpha^{-1}\vert \mathcal{B})} .
\end{displaymath}
\end{IEEEproof}

\subsection{Calculation of a Conditional Probability}
Denote by $\mathcal{B}_n=\sigma(Z_1, \cdots, Z_n)$ the observations obtained up to time $n$. Let $\mathcal{F}=\mathcal{F}_n$ and $\mathcal{B}=\mathcal{B}_n$. Then by Lemma 3.1, we have
\begin{equation}
P(A \vert \mathcal{B}_n)=\frac{E_Q(\alpha_n^{-1}\chi_A \vert \mathcal{B}_n)}{E_Q(\alpha_n^{-1}\vert \mathcal{B}_n)}~~(A \in \mathcal{F}_n) .
\end{equation}
In the previous section, $\alpha_n^{-1}$ has been calculated as
\begin{eqnarray}
\alpha_n^{-1} &=& \exp\left\{c\sum_{k=1}^n(X_k, Z_k)-\frac{1}{2}c^2n_0n \right\} \nonumber \\
&=& \beta_n . \nonumber
\end{eqnarray}
Note that $\beta_n$ is rewritten as
\begin{displaymath}
\beta_n=\exp\left\{\sum_{k=1}^n\left(\sum_{i=1}^{n_0}(cX_k^{(i)}Z_k^{(i)}-\frac{1}{2}c^2)\right)\right\} .
\end{displaymath}
\par
Here consider the probability density function of $z_j$ conditioned by $x_j$. This is given by
\begin{equation}
p(z_j \vert x_j)=\frac{1}{\sqrt{2 \pi}}\exp\left\{-\frac{(z_j-cx_j)^2}{2}\right\} .
\end{equation}
Hence, we have
\begin{eqnarray}
\log p(z_j \vert x_j) &=& K+\frac{1}{2}(2cx_jz_j-c^2(x_j)^2) \nonumber \\
&=& K+(cx_jz_j-\frac{1}{2}c^2) ,
\end{eqnarray}
where $K$ is a constant which depends only on $z_j$ and $(x_j)^2=1$ has been used. Accordingly,
\begin{equation}
\sum_{k=1}^n\left(\sum_{i=1}^{n_0}(cX_k^{(i)}Z_k^{(i)}-\frac{1}{2}c^2)\right)
\end{equation}
is just the {\it log-likelihood} function (i.e., metric)~\cite{lin 04} associated with the code sequence $\{X_k^{(i)}; 1 \leq i \leq n_0, 1 \leq k \leq n\}$. ($cX_k^{(i)}Z_k^{(i)}-\frac{1}{2}c^2$ is the symbol metric associated with the code symbol $X_k^{(i)}$ and $c(X_k, Z_k)-\frac{1}{2}c^2n_0$ is the branch metric.) Note that an exponential function having such a quantity as a power exponent can be equally regarded as a metric. Hence, $\beta_n$ is regarded as the metric associated with the code sequence $\{X_k^{(i)}; 1 \leq i \leq n_0, 1 \leq k \leq n\}$ as well. In the following, it is denoted by $\mbox{pm}(X)$. (Here a series of our arguments has been connected to convolutional coding/decoding.)
\par
Taking into consideration Property 3) of Proposition 2.1, it follows from Lemma 2.1 that
\begin{equation}
E_Q(\alpha_n^{-1}\vert \mathcal{B}_n)=\sum_{(\{-1,+1\}^{n_0})^n}\beta_n(x, Z_1, \cdots, Z_n)P_X(x) ,
\end{equation}
where $P_X(\cdot)$ is the distribution of $X_1, \cdots, X_n$. We remark that $P_X(\cdot)$ denotes the distribution with respect to the probability measure $Q$. However, from Property 1) of Proposition 2.1, this is identical to the distribution with respect to the original probability measure $P$. We also remark that $Z_1, \cdots, Z_n$ are mutually independent Gaussian random vectors with mean {\boldmath $0$} and covariance matrix $I_{n_0}$ from Property 2) of Proposition 2.1. This means the following. Let $\{z_k,~1 \leq k \leq n\}$ be a set of outcomes generated from the random vectors $Z_k=cX_k+W_k~(1 \leq k \leq n)$ under the probability measure $P$. Then we can equally regard $\{z_k,~1 \leq k \leq n\}$ as the outcomes of mutually independent Gaussian random vectors $Z_k~(1 \leq k \leq n)$ with mean {\boldmath $0$} and covariance matrix $I_{n_0}$ under the probability measure $Q$. That is, a set of outcomes can be seen in two different ways depending on the probability measures $P$ and $Q$.
\par
In the following, it is assumed that $k_0=1$ for simplicity. Let $C$ be a convolutional code generated by $G(D)$. Also, suppose that the corresponding code trellis is terminated in the all-zero state at depth $n$, without loss of generality. Hence, the number of effective information bits is $n-\nu$ ($\nu$ denotes the constraint length). Moreover, it is assumed that the information bits are equally likely. Under these assumptions, the occurrence probability of each code sequence $\{x_k,~1 \leq k \leq n\}$ is given by $\frac{1}{2^{n-\nu}}$. (Note that an actual encoded sequence is $\{\mbox{\boldmath $y$}_k,~1 \leq k \leq n\}$. However, there is a one-to-one correspondence between encoded sequences $\{\mbox{\boldmath $y$}_k,~1 \leq k \leq n\}$ and the corresponding code sequences $\{x_k,~1 \leq k \leq n\}~(x_k \in \{-1, +1\}^{n_0})$. Hence, we identify the latter with the former.)
\par
{\it Remark 1:} This paper is concerned with conditional probabilities of the form $P(\cdot \vert \mathcal{B}_n)$. Hence, it is natural to think of a maximum {\it a posteriori} probability (MAP) decoding algorithm (see Section III-F). In that case, the above metric is not appropriate. In fact, the assumption that the information bits are equally likely does not hold in general (for example, consider an iterative decoding algorithm~\cite{hage 96}). Let $S_l$ be a state at depth $l$ on the corresponding code trellis. In MAP decoding, the branch metric associated with a state transition $\eta_l=(S_{l-1}=s', S_l=s)$ (denoted by $\gamma_l(s', s)$) depends on the {\it a priori} probability of the associated information bit $i_l$. However, if the information bits are equally likely, then $\gamma_l(s', s)$ is essentially equal to the above (exponential) branch metric (see~\cite[Section 12.6]{lin 04}).
\par
{\it Remark 2:} As stated above, it is assumed that the code trellis is terminated in the all-zero state at depth $n$. On the other hand, we can consider a {\it truncated} convolutional code, where all ending states are possible at depth $n$. In this case, the conditional probability $P(X_n \in B \vert \mathcal{B}_n)$ corresponds to {\it filtering} of $X_n$ based on $\mathcal{B}_n$. Hence, in our situation, $P(X_l \in B \vert \mathcal{B}_n)~(l<n)$ corresponds to {\it smoothing} of $X_l$ based on $\mathcal{B}_n$.
\par
Now we have
\begin{eqnarray}
\lefteqn{E_Q(\alpha_n^{-1}\vert \mathcal{B}_n)} \nonumber \\
&& =\sum_{x \in C}\beta_n(x, Z_1, \cdots, Z_n)P_X(x) \nonumber \\
&& =\frac{1}{2^{n-\nu}}\sum_{x \in C}\mbox{pm}(x) .
\end{eqnarray}
\par
Next, note the relation
\begin{eqnarray}
\mathcal{F}_n &=& \sigma(X_k; k \in N)\vee \sigma(W_k; k \leq n) \nonumber \\
&=& \sigma(X_k; k \leq n)\vee \sigma(W_k; k \leq n) . \nonumber
\end{eqnarray}
For $1 \leq l \leq n$, $A=\{X_l \in B\}$ is contained in $\mathcal{F}_n$, where $B$ is a set of branch codes between depths $l-1$ and $l$. Then by Lemma 3.1, $E_Q(\alpha_n^{-1}\chi_A \vert \mathcal{B}_n)$ is calculated. That is, for $A=\{X_l \in B\}$, we obtain
\begin{eqnarray}
\lefteqn{E_Q(\alpha_n^{-1}\chi_A\vert \mathcal{B}_n)} \nonumber \\
&& =\sum_{x \in C, x_l \in B}\beta_n(x, Z_1, \cdots, Z_n)P_X(x) \nonumber \\
&& =\frac{1}{2^{n-\nu}}\sum_{x \in C, x_l \in B}\mbox{pm}(x) ,
\end{eqnarray}
where $\sum_{x \in C, x_l \in B}$ means that summation is carried out for those elements $x \in C$ such that $x_l \in B$.
\par
We finally have
\begin{eqnarray}
P(X_l \in B \vert \mathcal{B}_n) &=& \frac{\frac{1}{2^{n-\nu}}\sum_{x \in C, x_l \in B}\mbox{pm}(x)}{\frac{1}{2^{n-\nu}}\sum_{x \in C}\mbox{pm}(x)} \nonumber \\
&=& \frac{\sum_{x \in C, x_l \in B}\mbox{pm}(x)}{\sum_{x \in C}\mbox{pm}(x)} .
\end{eqnarray}
In particular, for $x_l=v$ ($v$ is a branch code), we have
\begin{equation}
P(X_l=v \vert \mathcal{B}_n)=\frac{\sum_{x \in C, x_l=v}\mbox{pm}(x)}{\sum_{x \in C}\mbox{pm}(x)} .
\end{equation}
Thus the following has been derived.
\begin{pro}
Let $C$ be a convolutional code generated by $G(D)$. It is assumed that the corresponding code trellis is terminated in the all-zero state at depth $n$. Let $B$ be a set of branch codes between depths $l-1$ and $l$ ($l \leq n$). Then the conditional probability $P(X_l \in B \vert \mathcal{B}_n)$ based on the observations up to time $n$ is given by
\begin{displaymath}
P(X_l \in B \vert \mathcal{B}_n)=\frac{\sum_{x \in C, x_l \in B}\mbox{pm}(x)}{\sum_{x \in C}\mbox{pm}(x)} .
\end{displaymath}
\end{pro}

\subsection{Conditional Probability in Recursive Form}
We know that when a convolutional code is represented using the associated code trellis, a state sequence $\{S_k\}$ has the Markov property~\cite{forn 731}. Hence, we see that a sequence of state transitions $\eta_k=(S_{k-1}, S_K)$ also has the Markov property. On the other hand, each state transition $\eta_k=(S_{k-1}, S_K)$ determines the associated branch code. Note that the converse is not true. However, when there is no danger of confusion, we identify a branch code $X_k$ with the associated state transition $\eta_k$. Noting these facts, the derived conditional probability $P(X_l \in B \vert \mathcal{B}_n)$ is transformed into a recursive form.
\par
Denote by $P_{k \vert k-1}(x_k \vert x_{k-1})$ the transition probability associated with a code sequence $\{X_k\}$. Also, let
\begin{equation}
b_k(x_k, z)\stackrel{\triangle}{=}\exp\{c(x_k, z)-\frac{1}{2}c^2n_0\} .
\end{equation}
Then $b_k(x_k, Z_k)$ represents the metric associated with the branch code $x_k$. Using the Markov property of $X_k~(k \in N)$, we have
\begin{eqnarray}
\lefteqn{P_X(x_1, \cdots, x_{l-1}, x_l, x_{l+1}, \cdots, x_n)} \nonumber \\
&& =P_X(x_1, \cdots, x_{l-1}) \times P_{l \vert l-1}(x_l \vert x_{l-1}) \times P_X(x_{l+1}, \cdots, x_n \vert x_l) .
\end{eqnarray}
Let us set $C_s^t\stackrel{\triangle}{=}\{(x_k, s \leq k \leq t); x \in C\}$. $C_s^t$ represents the set of code sub-sequences $(x_k, s \leq k \leq t)$, where each code sequence $x$ is restricted for the interval $s \leq k \leq t$. Then the numerator of $P(X_l \in B \vert \mathcal{B}_n)$ (denoted by $P(X_l \in B \vert \mathcal{B}_n)_{nu}$) is given by
\begin{eqnarray}
\lefteqn{P(X_l \in B \vert \mathcal{B}_n)_{nu}} \nonumber \\
&& =\sum_{x_l\in B}\left\{\sum_{x_1^{l-1} \in C_1^{l-1}}\beta_{l-1}(x_1, \cdots, x_{l-1}, Z_1, \cdots, Z_{l-1})P_X(x_1, \cdots, x_{l-1})\right \} \nonumber \\
&& \quad \times b_l(x_l, Z_l)P_{l \vert l-1}(x_l \vert x_{l-1})\left\{\sum_{x_{l+1}^n \in C_{l+1}^n}\beta_{l+1}^n(x_{l+1}^n, Z_{l+1}^n)P_X(x_{l+1}^n \vert x_l)\right\} .
\end{eqnarray}
Here, we have set
\begin{eqnarray}
\beta_s^t(x_s^t, Z_s^t) &\stackrel{\triangle}{=}& \beta_s^t(x_s, \cdots, x_t, Z_s, \cdots, Z_t) \\
P_X(x_s^t \vert x_{s-1}) &\stackrel{\triangle}{=}& P_X(x_s, \cdots, x_t \vert x_{s-1}) .
\end{eqnarray}
Accordingly, we have
\begin{eqnarray}
\beta_{l+1}^n(x_{l+1}^n, Z_{l+1}^n) &=& \beta_{l+1}^n(x_{l+1}, \cdots, x_n, Z_{l+1}, \cdots, Z_n) \nonumber \\
P_X(x_{l+1}^n \vert x_l) &=& P_X(x_{l+1}, \cdots, x_n \vert x_l) . \nonumber
\end{eqnarray}
\par
Similarly, the denominator of $P(X_l \in B \vert \mathcal{B}_n)$ (denoted by $P(X_l \in B \vert \mathcal{B}_n)_{de}$) is given by
\begin{eqnarray}
\lefteqn{P(X_l \in B \vert \mathcal{B}_n)_{de}} \nonumber \\
&& =\sum_{x_l\in C_l^l}\left\{\sum_{x_1^{l-1} \in C_1^{l-1}}\beta_{l-1}(x_1, \cdots, x_{l-1}, Z_1, \cdots, Z_{l-1})P_X(x_1, \cdots, x_{l-1})\right \} \nonumber \\
&& \quad \times b_l(x_l, Z_l)P_{l \vert l-1}(x_l \vert x_{l-1})\left\{\sum_{x_{l+1}^n \in C_{l+1}^n}\beta_{l+1}^n(x_{l+1}^n, Z_{l+1}^n)P_X(x_{l+1}^n \vert x_l)\right\} .
\end{eqnarray}
\par
As special cases, we have
\begin{eqnarray}
\lefteqn{P(X_1 \in B \vert \mathcal{B}_n)_{de}} \nonumber \\
&& =\sum_{x_1\in C_1^1}b_1(x_1, Z_1)P_X(x_1)\left\{\sum_{x_2^n \in C_2^n}\beta_2^n(x_2^n, Z_2^n)P_X(x_2^n \vert x_1)\right\}
\end{eqnarray}
\begin{eqnarray}
\lefteqn{P(X_n \in B \vert \mathcal{B}_n)_{de}} \nonumber \\
&& =\sum_{x_n\in C_n^n}\left \{\sum_{x_1^{n-1} \in C_1^{n-1}}\beta_{n-1}(x_1, \cdots, x_{n-1}, Z_1, \cdots, Z_{n-1})P_X(x_1, \cdots, x_{n-1})\right \} \nonumber \\
&& \quad \times b_n(x_n, Z_n)P_{n \vert n-1}(x_n \vert x_{n-1}) .
\end{eqnarray}
\par
Thus we have shown the following.
\begin{pro}
Under the same conditions as those for Proposition 3.1, the conditional probability $P(X_l \in B \vert \mathcal{B}_n)~(l \leq n)$ with respect to $\mathcal{B}_n$ is given by
\begin{equation}
P(X_l \in B \vert \mathcal{B}_n)=\frac{P(X_l \in B \vert \mathcal{B}_n)_{nu}}{P(X_l \in B \vert \mathcal{B}_n)_{de}} ,
\end{equation}
where the quantities on the right-hand side are defined as above.
\end{pro}

\subsection{Details of the Recursions}
We describe the recursions in more detail. Our final goal is to calculate the quantity
\begin{eqnarray}
\lefteqn{P(X_l \in B \vert \mathcal{B}_n)_{de}} \nonumber \\
&& =\sum_{x_l\in C_l^l}\left\{\sum_{x_1^{l-1} \in C_1^{l-1}}\beta_{l-1}(x_1, \cdots, x_{l-1}, Z_1, \cdots, Z_{l-1})P_X(x_1, \cdots, x_{l-1})\right \} \nonumber \\
&& \quad \times b_l(x_l, Z_l)P_{l \vert l-1}(x_l \vert x_{l-1})\left\{\sum_{x_{l+1}^n \in C_{l+1}^n}\beta_{l+1}^n(x_{l+1}^n, Z_{l+1}^n)P_X(x_{l+1}^n \vert x_l)\right\} . \nonumber
\end{eqnarray}
($P(X_l \in B \vert \mathcal{B}_n)_{nu}$ is calculated in a similar way.)
\par
\begin{figure}[htb]
\begin{center}
\includegraphics[width=8.0cm,clip]{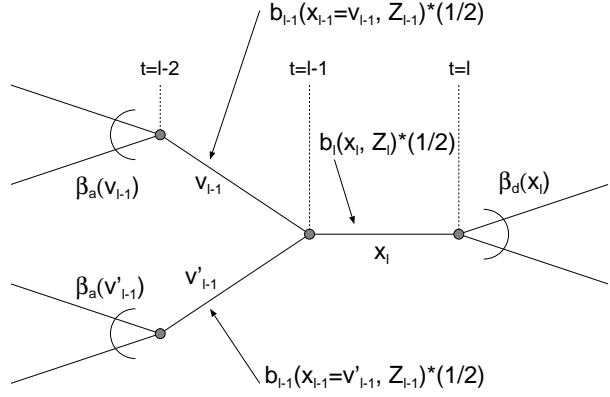}
\end{center}
\caption{Metrics associated with the code trellis.}
\label{Fig.1}
\end{figure}

\subsubsection{Forward Recursion}
First consider the forward recursion. Suppose that $\nu+2 \leq l$. When $x_l~(\in C_l^l)$ is fixed, the variable $x_{l-1}$ is restricted according to the transition probability $P_{l \vert l-1}(x_l \vert x_{l-1})$. Since $k_0=1$ is assumed, $P_{l \vert l-1}(x_l \vert x_{l-1})>0$ holds only for two values of $x_{l-1}$. Let these two values be $v_{l-1}$ and $v_{l-1}'$ (see Fig.1). Then we have
\begin{eqnarray}
\lefteqn{\left \{\sum_{x_1^{l-1} \in C_1^{l-1}}\beta_{l-1}(x_1, \cdots, x_{l-1}, Z_1, \cdots, Z_{l-1})P_X(x_1, \cdots, x_{l-1})\right \}\times P_{l \vert l-1}(x_l \vert x_{l-1})} \nonumber \\
&& =\sum_{x_1^{l-2}+v_{l-1} \in C_1^{l-1}}\beta_{l-1}(x_1^{l-2}, x_{l-1}=v_{l-1}, Z_1^{l-1})P_X(x_1^{l-2}, x_{l-1}=v_{l-1}) \times \frac{1}{2} \nonumber \\
&& \quad+\sum_{x_1^{l-2}+v_{l-1}' \in C_1^{l-1}}\beta_{l-1}(x_1^{l-2}, x_{l-1}=v_{l-1}', Z_1^{l-1})P_X(x_1^{l-2}, x_{l-1}=v_{l-1}') \times \frac{1}{2}  \nonumber \\
&& =\left \{\sum_{x_1^{l-2}+v_{l-1} \in C_1^{l-1}}\beta_{l-2}(x_1^{l-2}, Z_1^{l-2})P_X(x_1^{l-2})\right \} \times b_{l-1}(x_{l-1}=v_{l-1}, Z_{l-1}) \times \left(\frac{1}{2}\right)^2 \nonumber \\
&& \quad+\left \{\sum_{x_1^{l-2}+v_{l-1}' \in C_1^{l-1}}\beta_{l-2}(x_1^{l-2}, Z_1^{l-2})P_X(x_1^{l-2})\right \} \times b_{l-1}(x_{l-1}=v_{l-1}', Z_{l-1}) \times \left(\frac{1}{2}\right)^2 , \nonumber
\end{eqnarray}
where $x_1^{l-2}+v_{l-1}~(v_{l-1}')$ denotes the concatenation of $x_1^{l-2}$ and $v_{l-1}~(v_{l-1}')$. Also, we have used the relations:
\begin{eqnarray}
P_{l \vert l-1}(x_l \vert x_{l-1}=v_{l-1}) &=& P_{l \vert l-1}(x_l \vert x_{l-1}=v_{l-1}')=\frac{1}{2} \nonumber \\
P_{l-1 \vert l-2}(x_{l-1}=v_{l-1} \vert x_{l-2}) &=& P_{l-1 \vert l-2}(x_{l-1}=v_{l-1}' \vert x_{l-2})=\frac{1}{2} . \nonumber
\end{eqnarray}
Hence, we have
\begin{eqnarray}
\lefteqn{\left \{\sum_{x_1^{l-1} \in C_1^{l-1}}\beta_{l-1}(x_1, \cdots, x_{l-1}, Z_1, \cdots, Z_{l-1})P_X(x_1, \cdots, x_{l-1})\right \}} \nonumber \\
&& \quad \times b_l(x_l, Z_l)P_{l \vert l-1}(x_l \vert x_{l-1}) \nonumber \\
&& =\left \{\beta_a(v_{l-1}) \times b_{l-1}(x_{l-1}=v_{l-1}, Z_{l-1}) \times \frac{1}{2}+\beta_a(v_{l-1}') \times b_{l-1}(x_{l-1}=v_{l-1}', Z_{l-1}) \times \frac{1}{2}\right \} \nonumber \\
&& \quad \times \bigl(b_l(x_l, Z_l) \times \frac{1}{2}\bigr) ,
\end{eqnarray}
where
\begin{eqnarray}
\beta_a(v_{l-1}) &\stackrel{\triangle}{=}& \sum_{x_1^{l-2}+v_{l-1} \in C_1^{l-1}}\beta_{l-2}(x_1^{l-2}, Z_1^{l-2})P_X(x_1^{l-2}) \nonumber \\
\beta_a(v_{l-1}') &\stackrel{\triangle}{=}& \sum_{x_1^{l-2}+v_{l-1}' \in C_1^{l-1}}\beta_{l-2}(x_1^{l-2}, Z_1^{l-2})P_X(x_1^{l-2}) \nonumber .
\end{eqnarray}
We see that the above equation represents the forward recursion.
\par
The initial condition for the forward recursion is related to the initial ``transient'' sections of the corresponding code trellis. It is given by
\begin{equation}
\sum_{x_1 \in C_1^1}b_1(x_1, Z_1)P_X(x_1)=\sum_{x_1 \in C_1^1}\frac{1}{2} \times b_1(x_1, Z_1) .
\end{equation}
\par
Consider the next step in the forward recursion. We have
\begin{eqnarray}
\lefteqn{\sum_{x_1^2 \in C_1^2}\beta_2(x_1, x_2, Z_1, Z_2)P_X(x_1, x_2)} \nonumber \\
&& =\sum_{x_2 \in C_2^2}\left \{\sum_{x_1 \in C_1^1}b_1(x_1, Z_1)P_X(x_1)\right \} \times b_2(x_2, Z_2)P_{2 \vert 1}(x_2 \vert x_1) \nonumber \\
&& =\sum_{x_2 \in C_2^2}\left \{\sum_{x_1 \in C_1^1}\frac{1}{2} \times b_1(x_1, Z_1) \right \} \times b_2(x_2, Z_2)P_{2 \vert 1}(x_2 \vert x_1) . \nonumber
\end{eqnarray}
Note that $x_1$ is determined (denoted by $v_1$) given $x_2$. Hence, the right-hand side of the above equation becomes
\begin{displaymath}
\sum_{x_2 \in C_2^2}\left(\frac{1}{2}\right)^2 \times b_1(x_1=v_1, Z_1) \times b_2(x_2, Z_2) .
\end{displaymath}
\par
Continuing this procedure, we have
\begin{eqnarray}
\lefteqn{\sum_{x_1^l \in C_1^l}\beta_l(x_1, \cdots, x_l, Z_1, \cdots, Z_l)P_X(x_1, \cdots, x_l)} \nonumber \\
&& =\sum_{x_l \in C_l^l} \left(\frac{1}{2}\right)^l \times b_1(x_1=v_1, Z_1) \times b_2(x_2=v_2, Z_2) \times \cdots \times b_l(x_l, Z_l)
\end{eqnarray}
for $1 \leq l \leq \nu+1$.

\subsubsection{Backward Recursion}
Next, consider the backward recursion. This recursion is used to determine
\begin{displaymath}
\beta_d(x_l)\stackrel{\triangle}{=}\sum_{x_{l+1}^n \in C_{l+1}^n}\beta_{l+1}^n(x_{l+1}^n, Z_{l+1}^n)P_X(x_{l+1}^n \vert x_l) .
\end{displaymath}
Suppose that $l \leq n-\nu-1$. Then the backward recursion is expressed as
\begin{eqnarray}
\lefteqn{\sum_{x_{l+1}^n \in C_{l+1}^n}\beta_{l+1}^n(x_{l+1}^n, Z_{l+1}^n)P_X(x_{l+1}^n \vert x_l)} \nonumber \\
&& =\sum_{x_{l+1} \in C_{l+1}^{l+1}}b_{l+1}(x_{l+1}, Z_{l+1})P_{l+1 \vert l}(x_{l+1} \vert x_l) \nonumber \\
&& \quad \times \left\{\sum_{x_{l+2}^n \in C_{l+2}^n}\beta_{l+2}^n(x_{l+2}^n, Z_{l+2}^n)P_X(x_{l+2}^n \vert x_{l+1})\right\} \nonumber \\
&& =b_{l+1}(x_{l+1}=v_{l+1}, Z_{l+1})\times \frac{1}{2} \nonumber \\
&& \quad \times \left\{\sum_{x_{l+2}^n \in C_{l+2}^n}\beta_{l+2}^n(x_{l+2}^n, Z_{l+2}^n)P_X(x_{l+2}^n \vert x_{l+1}=v_{l+1})\right\} \nonumber \\
&& +b_{l+1}(x_{l+1}=v_{l+1}', Z_{l+1})\times \frac{1}{2} \nonumber \\
&& \quad \times \left\{\sum_{x_{l+2}^n \in C_{l+2}^n}\beta_{l+2}^n(x_{l+2}^n, Z_{l+2}^n)P_X(x_{l+2}^n \vert x_{l+1}=v_{l+1}')\right\} .
\end{eqnarray}
Here, we have used the relations that $P_{l+1 \vert l}(x_{l+1}=v_{l+1} \vert x_l)=\frac{1}{2}$ and $P_{l+1 \vert l}(x_{l+1}=v_{l+1}' \vert x_l)=\frac{1}{2}$.
\par
As in the forward recursion, the initial condition for the backward recursion is related to the final ``transient'' sections of the code trellis. It is given by
\begin{equation}
\sum_{x_n \in C_n^n}b_n(x_n, Z_n)P_{n \vert n-1}(x_n \vert x_{n-1})=b_n(x_n=v_n, Z_n) ,
\end{equation}
where $P_{n \vert n-1}(x_n=v_n \vert x_{n-1})=1$ is used. Note that this value depends on $x_{n-1}$. 
\par
Similarly, at the next step in the backward recursion, we have
\begin{eqnarray}
\lefteqn{\sum_{x_{n-1}^n \in C_{n-1}^n}\beta_{n-1}^n(x_{n-1}^n, Z_{n-1}^n)P_X(x_{n-1}^n \vert x_{n-2})} \nonumber \\
&& =\sum_{x_{n-1} \in C_{n-1}^{n-1}}b_{n-1}(x_{n-1}, Z_{n-1})P_{n-1 \vert n-2}(x_{n-1} \vert x_{n-2}) \times b_n(x_n=v_n, Z_n) \nonumber \\
&& =b_{n-1}(x_{n-1}=v_{n-1}, Z_{n-1})\times b_n(x_n=v_n, Z_n) , \nonumber
\end{eqnarray}
where $P_{n-1 \vert n-2}(x_{n-1}=v_{n-1} \vert x_{n-2})=1$ is used. Note that the value depends on $x_{n-2}$.
\par
Continuing this procedure, we have
\begin{eqnarray}
\lefteqn{\sum_{x_l^n \in C_l^n}\beta_l^n(x_l^n, Z_l^n)P_X(x_l^n \vert x_{l-1})} \nonumber \\
&& =b_l(x_l=v_l, Z_l)\times \cdots \times b_{n-1}(x_{n-1}=v_{n-1}, Z_{n-1}) \times b_n(x_n=v_n, Z_n)
\end{eqnarray}
for $n-\nu \leq l \leq n$. This value depends on $x_{l-1}$.

\subsection{Complexity of Calculating the Conditional Probability}
Let us evaluate the complexity required to calculate $P(X_l \in B \vert \mathcal{B}_{n+\nu})_{de}$. (A similar evaluation is possible for $P(X_l \in B \vert \mathcal{B}_n)_{nu}$.) It suffices to note the expression
\begin{eqnarray}
\lefteqn{\sum_{x_l\in C_l^l}\left\{\sum_{x_1^{l-1} \in C_1^{l-1}}\beta_{l-1}(x_1, \cdots, x_{l-1}, Z_1, \cdots, Z_{l-1})P_X(x_1, \cdots, x_{l-1})\right \}} \nonumber \\
&& \times b_l(x_l, Z_l)P_{l \vert l-1}(x_l \vert x_{l-1})\left\{\sum_{x_{l+1}^n \in C_{l+1}^n}\beta_{l+1}^n(x_{l+1}^n, Z_{l+1}^n)P_X(x_{l+1}^n \vert x_l)\right\} . \nonumber
\end{eqnarray}
\par
First fix $x_l~(\in C_l^l)$ arbitrarily and consider the summation
\begin{eqnarray}
\lefteqn{\sum_{x_1^{l-1} \in C_1^{l-1}}\beta_{l-1}(x_1, \cdots, x_{l-1}, Z_1, \cdots, Z_{l-1})P_X(x_1, \cdots, x_{l-1})} \nonumber \\
&& \qquad \qquad \times b_l(x_l, Z_l)P_{l \vert l-1}(x_l \vert x_{l-1}) . \nonumber
\end{eqnarray}
Since $x_l~(\in C_l^l)$ is fixed, the number of $x_1^{l-1}~(\in C_1^{l-1})$ is equal to that of code sub-sequences which start in {\boldmath $0$} (the all-zero state) at depth $0$ and end in state $S_{l-1}$ at depth $l-1$. It is given by $2^{l-1-\nu}$. Since $P_{l \vert l-1}(x_l \vert x_{l-1})>0$ holds only for two $x_{l-1}$'s given $x_l$ ($k_0$ is assumed to be $1$), the expression $2^{l-2-\nu} \times 2=2^{l-1-\nu}$ is more accurate (see the previous sub-section). Hence, the complexity of the summation is proportional to $2^{l-1-\nu}$.
\par
Next, consider the summation
\begin{displaymath}
\sum_{x_{l+1}^n \in C_{l+1}^n}\beta_{l+1}^n(x_{l+1}^n, Z_{l+1}^n)P_X(x_{l+1}^n \vert x_l) .
\end{displaymath}
The number of $x_{l+1}^n~(\in C_{l+1}^n)$ given $x_l$ is equal to that of code sub-sequences which start in state $S_l$ at depth $l$ and end in {\boldmath $0$} at depth $n$ and is given by $2^{n-\nu-l}$. Hence, the complexity of the summation is proportional to $2^{n-\nu-l}$.
\par
Finally, note that the number of $x_l~(\in C_l^l)$ is $2^{\nu+1}$.
\par
As a result, the total complexity $Q_c$ is given by
\begin{eqnarray}
Q_c &=& K_c \times 2^{l-1-\nu} \times 2^{n-\nu-l}\times 2^{\nu+1} \nonumber \\
&=& K_c \times 2^{n-\nu} ,
\end{eqnarray}
where $K_c$ is some constant. We remark that $2^{n-\nu}$ is the number of code sequences on the associated code trellis. Hence, the result is reasonable.

\subsection{Application to MAP Decoding}
Since we have obtained an expression for the conditional probability $P(X_l \in B \vert \mathcal{B}_n)$, we immediately see that it can be applied to MAP decoding~\cite{bahl 74,lin 04} of convolutional codes. Suppose that $k_0=1$ as before. Let {\boldmath $x$} and $\mbox{\boldmath $\hat x$}$ be the transmitted code sequence and the decoded code sequence, respectively. Here, taking into consideration the argument in Section III-B (see the remark in the sub-section), let us assume that the information bits are equally likely. When this assumption holds, the (ML) Viterbi algorithm maximizes $P(\mbox{\boldmath $\hat x$}=\mbox{\boldmath $x$} \vert \mathcal{B}_n)$. However, this does not guarantee that $P(\hat i_l=i_l \vert \mathcal{B}_n)$ is also maximized, where $i_l$ is the transmitted information bit and $\hat i_l$ is the decoded information bit (see~\cite[Section 12.6]{lin 04}). Hence, although the above assumption restricts the applications, we can still consider MAP decoding.
\par
Let $i_l$ be the information bit at $t=l$ and suppose that a state at depth $l$ on the code trellis has the form $S_l=(i_{l-\nu+1}, \cdots, i_{l-1}, i_l)$. Denote by $B_0$ the set of branch codes whose branches enter into the states $S_l^0=(\cdots, i_l=0)$ at depth $l$. Similarly, denote by $B_1$ the set of branch codes whose branches enter into the states $S_l^1=(\cdots, i_l=1)$ at depth $l$. Then we have
\begin{eqnarray}
P(i_l=0 \vert \mathcal{B}_n) &=& P(X_l \in B_0 \vert \mathcal{B}_n) \nonumber \\
P(i_l=1 \vert \mathcal{B}_n) &=& P(X_l \in B_1 \vert \mathcal{B}_n) . \nonumber
\end{eqnarray}
Using these equations, the ratio of the {\it a posteriori} probability (APP) of $i_l$ being $0$ to the APP of $i_l$ being $1$ is calculated as
\begin{equation}
\Lambda(i_l)=\frac{P(i_l=0 \vert \mathcal{B}_n)}{P(i_l=1 \vert \mathcal{B}_n)}~~(1 \leq l \leq n).
\end{equation}
Hence, MAP decoding of convolutional codes can be realized based on the above APP ratio $\Lambda(i_l)$.

\section{Conclusion}
We have considered a standard observation model where a convolutionally encoded sequence is transmitted symbol by symbol over an AWGN channel using BPSK modulation and have discussed it from the viewpoint of filtering (smoothing) for discrete-time stochastic processes. In this case, since pairs of the signal and observation are not jointly Gaussian, a linear estimation method cannot be used. Then we have applied a non-linear estimation method to the problem. (We have used the argument given in~\cite{kuni 76}.) More precisely, we have used a discrete-time version of the {\it Girsanov theory}, which states a finite-dimensional Gaussian distribution is invariant under appropriate shifts of variables and a transformation of the underlying probability measure. As a result, we have derived the conditional probability of an event related to any encoded symbol conditioned by the observations. We have also transformed it into a recursive form. Moreover, we have shown that the derived conditional probability can be used for MAP decoding of convolutional codes. We remark that the method in this paper can be applied to block codes as well. We think a connection between the coding theory and the estimation theory for stochastic processes has been more clarified through the discussion.


%

\appendices
\section{Proof of Lemma 2.1}
\renewcommand{\theequation}{\thesection.\arabic{equation}}
\addtocounter{equation}{-51}
Let $\mathcal{B}_{d_1}$ and $\mathcal{B}_{d_2}$ be the Borel $\sigma$-fields in $R^{d_1}$ and $R^{d_2}$, respectively. If $u(x, y)$ has the form $u(x, y)=u_1(x)u_2(y)$, then the lemma holds. Consider a general case. Let $\{A_j \times B_j,~1 \leq j \leq j_0\}$ be pairwise disjoint, where $A_j \in \mathcal{B}_{d_1}$ and $B_j \in \mathcal{B}_{d_2}$. Denote by $\sum_{j=1}^{j_0}A_j \times B_j$ the union of $A_j \times B_j~(1 \leq j \leq j_0)$. In this case, we have
\begin{eqnarray}
\chi_{\sum_{j=1}^{j_0}A_j \times B_j}(x, y) &=& \sum_{j=1}^{j_0}\chi_{A_j \times B_j}(x, y) \nonumber \\
&=& \sum_{j=1}^{j_0}\chi_{A_j}(x) \times \chi_{B_j}(y) . \nonumber
\end{eqnarray}
Note that a family of sets of the form $\sum_{j=1}^{j_0}A_j \times B_j$ generates the $\sigma$-field $\mathcal{B}_{d_1} \times \mathcal{B}_{d_2}$ in $R^{d_1+d_2}$~\cite{hew 65,nev 65}. Using these facts, it is shown that $u(x, y)$ is approximated by a linear combination of functions of the form $u_1(x)u_2(y)$.

\section{Proof of Proposition 2.1}
Since $\alpha_0=1$, $Q(A)=\int_A\alpha_0dP=P(A)$ holds for $A \in \mathcal{F}_0$. Here note the relation $\mathcal{F}_0=\sigma(X_k; k \in N)$. Thus 1) is proved.
\par
Next, let us show 2). For $n \times n_0$ vector $\xi=(\xi_1, \cdots, \xi_n)$, we define the following quantity:
\begin{equation}
\gamma_n^{\xi}\stackrel{\triangle}{=}\exp\left\{i\sum_{k=1}^n(\xi_k, W_k+cX_k)+\frac{1}{2}\sum_{k=1}^n \vert \xi_k \vert^2 \right\}~~(\gamma_0^{\xi}=1) ,
\end{equation}
where $i\stackrel{\triangle}{=}\sqrt{-1}$. Then we have
\begin{equation}
\alpha_n \times \gamma_n^{\xi}=\exp\left\{\sum_{k=1}^n(i \xi_k-cX_k, W_k)-\frac{1}{2}\sum_{k=1}^n \vert i \xi_k-cX_k \vert^2 \right\} .
\end{equation}
By Lemma 2.1,
\begin{eqnarray}
\lefteqn{E(\exp\{(i\xi_n-cX_n, W_n)-\frac{1}{2}\vert i\xi_n-cX_n \vert^2\} \vert \mathcal{F}_{n-1})} \nonumber \\
&=& \frac{1}{(2 \pi)^{n_0/2}}\int_{R^{n_0}}\exp\{(i\xi_n-cX_n, y)-\frac{1}{2}\vert i\xi_n-cX_n \vert^2\}\exp\{-\frac{1}{2}(y, y)\}dy \nonumber \\
&=& \frac{1}{(2 \pi)^{n_0/2}}\int_{R^{n_0}}\exp\{-\frac{1}{2}(y-u_n, y-u_n)\}dy=1
\end{eqnarray}
is obtained, where we have set $u_n=i\xi_n-cX_n$. (The last equality is derived by applying the {\it Cauchy integral theorem} in complex analysis.)
\par
Using this fact, it is shown that $\{\alpha_n \gamma_n^{\xi}\}$ is an $\mbox{\boldmath $\mathcal{F}$}$-martingale. A proof is similar to that for $\alpha_n$. Hence, for $m<n~(\leq n_p)$,
\begin{displaymath}
E(\alpha_n \gamma_n^{\xi} \vert \mathcal{F}_m)=\alpha_m \gamma_m^{\xi}
\end{displaymath}
holds and for $A \in \mathcal{F}_m$, we have
\begin{displaymath}
\int_AE(\alpha_n \gamma_n^{\xi} \vert \mathcal{F}_m)dP=\int_A\alpha_m \gamma_m^{\xi}dP .
\end{displaymath}
The left-hand side is equal to $\int_A\alpha_n \gamma_n^{\xi}dP$ from the definition of a conditional expectation. Hence,
\begin{equation}
\int_A\alpha_n \gamma_n^{\xi}dP=\int_A\alpha_m \gamma_m^{\xi}dP~~(A \in \mathcal{F}_m)
\end{equation}
is obtained.
\par
On the other hand, if $A \in \mathcal{F}_m$, then $A \in \mathcal{F}_n$ and by Lemma 2.3, we have
\begin{equation}
Q(A)=\int_A\alpha_n dP=\int_A\alpha_m dP .
\end{equation}
Hence, both $dQ=\alpha_n dP$ and $dQ=\alpha_m dP$ hold on the $\sigma$-field $\mathcal{F}_m$ and it follows that
\begin{displaymath}
 \int_A\gamma_n^{\xi}dQ=\int_A\alpha_n \gamma_n^{\xi}dP=\int_A\alpha_m \gamma_m^{\xi}dP=\int_A\gamma_m^{\xi}dQ .
\end{displaymath}
That is, for $A \in \mathcal{F}_m$, we have
\begin{equation}
\int_A\gamma_n^{\xi}dQ=\int_A\gamma_m^{\xi}dQ .
\end{equation}
This implies that
\begin{displaymath}
\gamma_n^{\xi}dQ=\gamma_m^{\xi}dQ
\end{displaymath}
or equivalently,
\begin{displaymath}
\gamma_n^{\xi}(\gamma_m^{\xi})^{-1}dQ=dQ
\end{displaymath}
on the $\sigma$-field $\mathcal{F}_m$. Then
\begin{equation}
\int_A\gamma_n^{\xi}(\gamma_m^{\xi})^{-1}dQ=\int_AdQ=Q(A)~~(m<n)
\end{equation}
is obtained for $A \in \mathcal{F}_m$. Note that this is rewritten as
\begin{equation}
\int_A\exp\left\{i\sum_{k=m+1}^n(\xi_k, W_k+cX_k)\right\}dQ=\exp\left\{-\frac{1}{2}\sum_{k=m+1}^n \vert \xi_k \vert^2 \right\}\times Q(A)~~(m<n) .
\end{equation}
Here note the relation $W_k+cX_k=Z_k$. By letting $A=\Omega$ and $m=0$, 2) is proved from the property of a characteristic function~\cite{ito 91,jaz 07,kuni 76}.
\par
Finally, let us show 3). Again, note the relation
\begin{displaymath}
\int_A\exp\left\{i\sum_{k=m+1}^n(\xi_k, Z_k)\right\}dQ=\exp\left\{-\frac{1}{2}\sum_{k=m+1}^n \vert \xi_k \vert^2 \right\}\times Q(A)~~(m<n) .
\end{displaymath}
It is modified as
\begin{displaymath}
\int \chi_A \times \exp\left\{i\sum_{k=m+1}^n(\xi_k, Z_k)\right\}dQ=\int \chi_AdQ \int \exp\left\{i\sum_{k=m+1}^n(\xi_k, Z_k)\right\}dQ ,
\end{displaymath}
where $\chi_A$ the indicator function of $A$. Hence, for a simple function~\cite{hew 65}
\begin{displaymath}
\sum_{p=1}^{p_0}a_p\chi_{A_p}~~(a_p \in R,~A_p \in \mathcal{F}_m) ,
\end{displaymath}
we have
\begin{eqnarray}
\lefteqn{\int \left(\sum_{p=1}^{p_0}a_p\chi_{A_p}\right) \times \exp\left\{i\sum_{k=m+1}^n(\xi_k, Z_k)\right\}dQ} \nonumber \\
&& =\int \left(\sum_{p=1}^{p_0}a_p\chi_{A_p}\right)dQ \int \exp\left\{i\sum_{k=m+1}^n(\xi_k, Z_k)\right\}dQ .
\end{eqnarray}
Let $U$ be an arbitrary $\mathcal{F}_m$-measurable function. In the following, we will show that
\begin{eqnarray}
\lefteqn{\int \exp\{i\xi U\} \times \exp\left\{i\sum_{k=m+1}^n(\xi_k, Z_k)\right\}dQ} \nonumber \\
&& =\int \exp\{i\xi U\}dQ \int \exp\left\{i\sum_{k=m+1}^n(\xi_k, Z_k)\right\}dQ
\end{eqnarray}
holds.
\par
In the expression
\begin{displaymath}
\exp\{i\xi U\}=\cos(\xi U)+i\sin(\xi U) ,
\end{displaymath}
both $\cos(\xi U)$ and $\sin(\xi U)$ are $\mathcal{F}_m$-measurable and can be approximated by simple functions~\cite{hew 65,nev 65} as
\begin{eqnarray}
\cos(\xi U) &\approx& \sum_{p=1}^{p_0}a_p\chi_{A_p}~~(a_p \in R,~A_p \in \mathcal{F}_m) \\
\sin(\xi U) &\approx& \sum_{q=1}^{q_0}b_q\chi_{B_q}~~(b_q \in R,~B_q \in \mathcal{F}_m) .
\end{eqnarray}
Here note that
\begin{eqnarray}
\lefteqn{\int \left(\sum_{q=1}^{q_0}b_q\chi_{B_q}\right) \times \exp\left\{i\sum_{k=m+1}^n(\xi_k, Z_k)\right\}dQ} \nonumber \\
&& =\int \left(\sum_{q=1}^{q_0}b_q\chi_{B_q}\right)dQ \int \exp\left\{i\sum_{k=m+1}^n(\xi_k, Z_k)\right\}dQ
\end{eqnarray}
also holds. Hence, we have
\begin{eqnarray}
\lefteqn{\int \left(\sum_{p=1}^{p_0}a_p\chi_{A_p}\right) \times \exp\left\{i\sum_{k=m+1}^n(\xi_k, Z_k)\right\}dQ} \nonumber \\
&& \quad+i\int \left(\sum_{q=1}^{q_0}b_q\chi_{B_q}\right) \times \exp\left\{i\sum_{k=m+1}^n(\xi_k, Z_k)\right\}dQ \nonumber \\
&& =\int \left(\sum_{p=1}^{p_0}a_p\chi_{A_p}\right)dQ \int \exp\left\{i\sum_{k=m+1}^n(\xi_k, Z_k)\right\}dQ \nonumber \\
&& \quad+i\int \left(\sum_{q=1}^{q_0}b_q\chi_{B_q}\right)dQ \int \exp\left\{i\sum_{k=m+1}^n(\xi_k, Z_k)\right\}dQ \nonumber \\
&& =\left(\int \left(\sum_{p=1}^{p_0}a_p\chi_{A_p}\right)dQ+i\int \left(\sum_{q=1}^{q_0}b_q\chi_{B_q}\right)dQ\right) \nonumber \\
&& \quad \times \int \exp\left\{i\sum_{k=m+1}^n(\xi_k, Z_k)\right\}dQ . \nonumber
\end{eqnarray}
Thus an equality
\begin{eqnarray}
\lefteqn{\int \left(\sum_{p=1}^{p_0}a_p\chi_{A_p}+i\sum_{q=1}^{q_0}b_q\chi_{B_q}\right) \times \exp\left\{i\sum_{k=m+1}^n(\xi_k, Z_k)\right\}dQ} \nonumber \\
&& =\int \left(\sum_{p=1}^{p_0}a_p\chi_{A_p}+i\sum_{q=1}^{q_0}b_q\chi_{B_q}\right)dQ \int \exp\left\{i\sum_{k=m+1}^n(\xi_k, Z_k)\right\}dQ
\end{eqnarray}
has been shown. Since $\exp\{i\xi U\}$ can be approximated by simple functions $\sum_{p=1}^{p_0}a_p\chi_{A_p}+i\sum_{q=1}^{q_0}b_q\chi_{B_q}$ with arbitrary accuracy~\cite{hew 65,nev 65}, the above equality means that
\begin{eqnarray}
\lefteqn{\int \exp\{i\xi U\} \times \exp\left\{i\sum_{k=m+1}^n(\xi_k, Z_k)\right\}dQ} \nonumber \\
&& =\int \exp\{i\xi U\}dQ \int \exp\left\{i\sum_{k=m+1}^n(\xi_k, Z_k)\right\}dQ \nonumber
\end{eqnarray}
holds. This final expression implies that $Z_k~(m+1 \leq k \leq n)$ and $\mathcal{F}_m$ are mutually independent~\cite{jaz 07}. In particular, letting $m=0$ and $n=n_p$, we have 3).




\ifCLASSOPTIONcaptionsoff
  \newpage
\fi


\begin{thebibliography}{99}
\bibitem{ari 77}S.~Arimoto, {\em Kalman Filter}, (in Japanese). Tokyo, Japan: Sangyo Tosho Publishing, 1977.
\bibitem{bahl 74}L.~R.~Bahl, J.~Cocke, F.~Jelinek, and J.~Raviv, ``Optimal decoding of linear codes for minimizing symbol error rate,'' {\em IEEE Trans.~Inf.~Theory}, vol.~IT-20, no.~2, pp.~284--287, March 1974.
\bibitem{bala 73}A.~V.~Balakrishnan, {\em Stochastic Differential Systems I}, Lecture Notes in Economics and Mathematical Systems, 84. Springer-Verlag Berlin Heidelberg New York, 1973.
\bibitem{forn 70}G.~D.~Forney, Jr., ``Convolutional codes I: Algebraic structure,'' {\em IEEE Trans.~Inf.~Theory}, vol.~IT-16, no.~6, pp.~720--738, Nov. 1970.
\bibitem{forn 731}G.~D.~Forney, Jr., ``The Viterbi algorithm,'' {\em Proc. IEEE}, vol.~61, no.~3, pp.~268--278, March 1973.
\bibitem{fuji 72}M.~Fujisaki, G.~Kallianpur, and H.~Kunita, ``Stochastic differential equations for the non linear filtering problem,'' {\em Osaka J. Math.}, vol.~9, no.~1, pp.~19--40, 1972.
\bibitem{gir 60}I.~V.~Girsanov, ``On transforming a certain class of stochastic processes by absolutely continuous substitution of measures,'' (English transl.), {\em Theory of Prob. and Appl.}, vol.~5, no.~3, pp.~285--301, 1960.
\bibitem{hage 96}J.~Hagenauer, E.~Offer, and L.~Papke, ``Iterative decoding of binary block and convolutional codes,'' {\em IEEE Trans.~Inf.~Theory}, vol.~42, no.~2, pp.~429--445, March 1996.
\bibitem{hell 71}J.~A.~Heller and I.~M.~Jacobs, ``Viterbi decoding for satellite and space communication,'' {\em IEEE Trans.~Commun.~Technol.}, vol.~COM-19, no.~5, pp.~835--848, Oct. 1971.
\bibitem{hew 65}E.~Hewitt and K.~Stromberg, {\em Real and Abstract Analysis}. Springer-Verlag Berlin Heidelberg, 1965.
\bibitem{hida 80}T.~Hida, {\em Brownian Motion}, (English transl.). New York, USA: Springer-Verlag, 1980.
\bibitem{ito 91}K.~Ito, {\em Probability Theory}, (in Japanese). Tokyo, Japan: Iwanami Shoten, 1991.
\bibitem{jaz 07}A.~H.~Jazwinski, {\em Stochastic Processes and Filtering Theory}. New York, USA: Dover Publications, 2007.
\bibitem{joha 99}R.~Johannesson and K.~S.~Zigangirov, {\em Fundamentals of Convolutional Coding}. New York, USA: IEEE Press, 1999.
\bibitem{kai 681}T.~Kailath, ``An innovations approach to least-squares estimation--Part I: Linear filtering in additive white noise,'' {\em IEEE Trans.~Automatic Control}, vol.~AC-13, no.~6, pp.~646--655, Dec. 1968.
\bibitem{kai 682}T.~Kailath and P.~Frost, ``An innovations approach to least-squares estimation--Part II: Linear smoothing in additive white noise,'' {\em IEEE Trans.~Automatic Control}, vol.~AC-13, no.~6, pp.~655--660, Dec. 1968.
\bibitem{kai 702}T.~Kailath, ``A further note on a general likelihood formula for random signals in Gaussian noise,'' {\em IEEE Trans.~Inf.~Theory}, vol.~IT-16, no.~4, pp.~393--396, July 1970.
\bibitem{kai 98}T.~Kailath and H.~V.~Poor, ``Detection of stochastic processes,'' {\em IEEE Trans.~Inf.~Theory}, vol.~44, no.~6, pp.~2230--2259, Oct. 1998.
\bibitem{kara 91}I.~Karatzas and S.~E.~Shreve, {\em Brownian Motion and Stochastic Calculus}, 2nd ed. Springer-Verlag New York, 1991.
\bibitem{kuni 76}H.~Kunita, {\em Estimation of Stochastic Processes}, (in Japanese). Tokyo, Japan: Sangyo Tosho Publishing, 1976.
\bibitem{lin 04}S.~Lin and D.~J.~Costello, Jr., {\em Error Control Coding}, 2nd ed. Upper Saddle River, NJ, USA: Prentice-Hall, 2004.
\bibitem{nev 65}J.~Neveu, {\em Mathematical Foundations of the Calculus of Probability}, (English transl.). San Francisco, USA: Holden-Day, 1965.
\bibitem{oksen 98}B.~{\O}ksendal, {\em Stochastic Differential Equations--An Introduction with Applications}, 5th ed. Springer-Verlag Berlin Heidelberg, 1998.
\bibitem{taji 19}M.~Tajima, ``An innovations approach to Viterbi decoding of convolutional codes,'' {\em IEEE Trans.~Inf.~Theory}, vol.~65, no.~5, pp.~2704--2722, May 2019.
\bibitem{will 91}D.~Williams, {\em Probability with Martingales}. Cambridge University Press, 1991. (J.~Akahori et al., {\em Probability with Martingales}, (Japanese transl.). Tokyo, Japan: Baifukan, 2004.)
\bibitem{wong 71}E.~Wong, {\em Stochastic Processes in Information and Dynamical Systems}. New York, USA: McGraw-Hill, 1971.
\bibitem{wong 73}E.~Wong, ``Recent progress in stochastic processes--A survey,'' {\em IEEE Trans.~Inf.~Theory}, vol.~IT-19, no.~3, pp.~262--275, May 1973.
\end{thebibliography}
\end{document}